\documentclass[12pt]{article}

\usepackage{amsmath,amsfonts,amssymb,latexsym,graphicx}

\usepackage{caption}
\usepackage{subcaption}

\def\be{\begin{equation}}
\def\ee{\end{equation}}
\def\bq{\begin{eqnarray}}
\def\eq{\end{eqnarray}}
\def\beq{\begin{eqnarray}}
\def\eeq{\end{eqnarray}}

\def\pa{\partial}

\begin{document}
%\pagenumbering{gobble}
\Large
\begin{center}
\textsc{Legendre scalarization in gravity and cosmology}
\end{center}
\begin{center}
\Large
Spiros Cotsakis$^{1,2,\dagger}$, Jose P. Mimoso$^{3,\ddagger}$, John Miritzis$^{4,*}$
\end{center}
\begin{center}
\small
$^1$) Institute of Gravitation and Cosmology, RUDN University\\
ul. Miklukho-Maklaya 6, Moscow 117198, Russia\\
$^{2}$) Research Laboratory of Geometry,  Dynamical Systems\\  and Cosmology,
University of the Aegean,\\ Karlovassi 83200,  Samos, Greece\\
${\dagger} $skot@aegean.gr\\
$^{3}$) Departamento de F\'{\i}sica and Instituto de Astrof\'{\i}sica e Ci\^encias do Espa\c co\\
Faculdade de Ci\^encias, Universidade de Lisboa\\
Ed. C8, Campo Grande, 1769-016 Lisboa, Portugal\\
${\ddagger}$ jpmimoso@fc.ul.pt\\
$^{4}$) Department of Marine Sciences\\University of the Aegean\\University Hill, Mytilene 81100, Greece\\
${*}$ imyr@aegean.gr
\end{center}

\begin{center}\small May 2023\end{center}
\vspace{3pt}
\normalsize

\noindent \textbf{Abstract.}  We propose a new formulation of $f(R)$ gravity, dubbed scalarized $f(R)$ gravity, in which the Legendre transform is included as a dynamical term. This leads to a theory with second-order field equations  that describes general relativity with a self-interacting scalar field, without requiring the introduction of conformal frames. We demonstrate that the quadratic version of scalarized $f(R)$ gravity reduces to general relativity with a massive scalar field, and we explore its implications for  Friedmann cosmology. Our findings suggest that scalarized $f(R)$ gravity may lead to simplified descriptions of cosmological applications, while the proposed formulation could offer a new perspective on the relationship between $f(R)$ gravity and scalar-tensor theories.

\section{Introduction}
It is well-known that the conformal transformation approach between Jordan and Einstein frames recasts vacuum modified gravity, be it scalar-tensor or higher-order, in a useful, reduced-order form  as general relativity plus a self-interacting scalar field with a particular conformal potential (also true in the case of matter fields present in both frames with additional matter-scalar field couplings), cf. \cite{d2}-\cite{cy2}. By a slightly more complicated but totally equivalent procedure, one may also first introduce the Legendre transform of higher-order gravity to express  higher-order gravity in a scalar-tensor form with the Legendre transform of the theory regarded as potential term, and then   conformally transformed the resulting scalar-tensor theory to obtain the same result as before in the Einstein frame representation of the original lagrangian \cite{ha}-\cite{jk}.

In the ensuing decades since the first conformal equivalence results originally appeared, we have learnt that two conformally related theories can have a number of common physical properties that arise in a great variety of different problems in gravity and cosmology (inflation, stability, black holes, quantum aspects, etc), as can be seen in the vast literature on this subject (cf.  the quoted papers above and refs. therein), and play a major role in providing models for the interpretation of recent cosmological observations \cite{amendola}-\cite{riess}.  However, two conformally related metrics satisfying different field equations are generally  physically inequivalent because their rescaling is not constant, in sharp contrast to  them enjoying  common physical  properties. How can one reconcile the two?

In this paper, we introduce a procedure, which we call `Legendre scalarization', that has the feature of turning an $f(R)$ gravity theory into general relativity with a self-interacting scalar field while keeping the same spacetime metric. In the next Section, we show that in the  resulting  Legendre-scalarized theory, the Legendre transform of the original theory is combined with the kinetic term of the scalar field involved in it, to become dynamical terms in a new theory which we study in detail. We  show that the scalarized version of the original higher-order lagrangian (in vacuum or with matter)  describes general relativity with the Legendre scalar field playing the role of a self-interacting scalar field with potential given by the Legendre transform of the theory, but in the same metric of the original lagrangian. In Section 3, we apply this procedure to a homogeneous and isotropic  Friedmann universe and study the resulting equations in the case of a theory with a quadratic lagrangian in the scalar curvature. In the last Section, we briefly discuss these results and possible extensions of the mechanism introduced here.

\section{Einstein-Legendre gravity}
\begin{figure}[tbh]
\begin{center}
\includegraphics{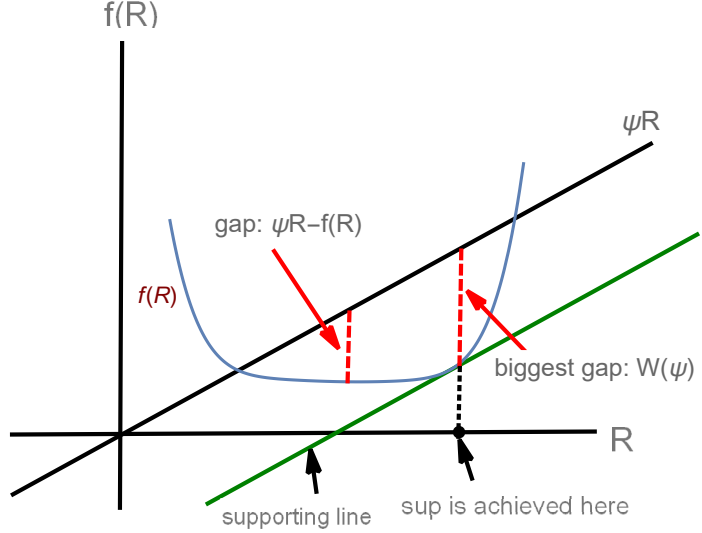}
\end{center}
\par
\caption{The Legendre transform $W(\psi)$ of $f(R)$.}%
\label{leg}%
\end{figure}

We start with a smooth $f(R)$ function of the scalar curvature $R$ and  assume that $f''(R)$ exists and has constant sign. %We consider the Legendre transform of the function $f(R)$ defined as follows.
In the $(R$-$f(R))$-plane, the lagrangian of vacuum general relativity is the bisector of the first and third quadrants. For the $f(R)$ curve, let $\psi$ be its slope at a given point, $\psi=f'(R)$, and consider the line $\psi R$ passing through the origin, cf. Fig. \ref{leg}. The difference $\psi R -f(R)$ describes the gap between the line $\psi R$ and the curve $f(R)$, and this gap becomes biggest at the point where we can shift the line $\psi R$ to be the tangent  `supporting' line to  the curve $f(R)$.
The supporting line is expressed by the \emph{Legendre (or $W$-) transform} $W(\psi)$ of $f(R)$ which is defined to be \cite{ar}, p.64, \cite{zia},
\be \label{let}
W(\psi)=F(\psi,R(\psi))=\sup_{R\in\mathbb{R}}\left\{F(\psi,R)\right\},
\ee
where we have set $F(\psi,R)=\psi R -f(R)$, and $\psi$ is the slope $\partial f/\partial R$ of $f$ like before. When $R=R(\psi)$, in practice this means that $R=(f^{-1}(\psi))$, and we get the function $W(\psi)$ that depends only of $\psi$, with $W(\psi)=F(\psi,R(\psi))$. The function $W(\psi)$ encodes the information contained in $f(R)$ near the point $R$ where the supremum is attained, and describes how much we have to shift the line $\psi R$ at each point to become the supporting line of $f(R)$.

Therefore for fixed $\psi$, the Legendre transform $W(\psi)$ of $f(R)$ is linear and becomes the envelope of $f(R)$, that is the bundle of all supporting lines of a given $f(R)$ constructed this way.
When the Legendre transform $W(\psi)$ of $f(R)$ exists, we can write  $f(R)=\psi R-W(\psi)$, and this is sometimes used in the literature of modified gravity to express the $f(R)$ theory as a Brans-Dicke theory with `potential' $W(\psi)$  but without a kinetic term (this corresponds to zero Brans-Dicke parameter $\omega$), cf. \cite{ha}-\cite{jk}.

It follows from (\ref{let}) that $F(\psi,R)\leq W(\psi)$, and this leads to  Young's inequality,   $\psi R\leq f(R) +W(\psi)$. For example, when $f(R)=R^2/2$, we have $W(\psi)=\psi^2/2$, and so we find that $\psi R\leq R^2/2+\psi^2/2$. Another basic property that follows from the Legendre transform is that the  equations obtained by varying the action associated with the Lagrangian $F(\psi,R)=\psi R -f(R)$ with respect to the spacetime metric $g$ are equivalent to the lagrangian  equations obtained by varying the action associated with the lagrangian $f(R)$.

For example, let us take $f_2(R)=R+\epsilon R^2$, and consider  the associated $W_2(\psi)$-action $\int_\mathcal{M} W_2(\psi)\,dv_g$, where $dv_g$ denotes the volume element of the spacetime manifold $\mathcal{M}$ with respect to the metric $g_{\mu\nu}$ (in local coordinates $x^\mu$, $dv_g=\sqrt{g}\,dx^\mu$, with $g=\textrm{det} g_{\mu\nu}$). If we vary this action with respect to the $\psi$-family of fields $\{\psi(s):s\in\mathbb{R}\}$, by setting $\dot{\psi}=(\partial \psi/\partial s)|_{s=0}$, with $\psi(0)=\psi$, it follows that this fixes $\psi=1$, and we get $\epsilon=0$ in the expression of $f_2(R)$, that is general relativity (the $dv_g$-variation with respect to $\psi$ is zero, $\dot{R}=(\pa R/\pa\psi)\dot{\psi}$, etc).

This suggests that the general \emph{metric} (as opposed to the $\psi$-) variation of the  `$W$-action',
\be\label{F}
S_W=\int_\mathcal{M} F(\psi,R)\,dv_g=\int_\mathcal{M} (\psi R-f(R))\,dv_g,
\ee
is somehow related to the Einstein-Hilbert lagrangian.  Since,  $S_W=S_{BD}-S_{f(R)}$, with $S_{BD}=\int_\mathcal{M} \psi R\,dv_g$  the geometric part of the standard Brans-Dicke action, and $S_{f(R)}=\int_\mathcal{M} f(R)\,dv_g$ the standard vacuum $f(R)$ action, we consider the family of metrics $\{g_{\mu\nu}(s):s\in\mathbb{R}\}$, and write $\dot{g}_{\mu\nu}=(\partial g_{\mu\nu}/\partial s)|_{s=0}$, for the metric variation, with $g_{\mu\nu}(0)=g_{\mu\nu}$, and we find that,
\be\label{varBD}
\dot{S}_{BD}=\int_\mathcal{M}dv_g\, \dot{g}^{\mu\nu}\,\left(\psi\,(R_{\mu\nu}-\frac{1}{2}g_{\mu\nu}R)-\nabla_\mu\nabla_\nu\psi+g_{\mu\nu}\Box\psi
\right),
\ee
and,
\be\label{varfofR}
\dot{S}_{f(R)}=\int_\mathcal{M}dv_g\, \dot{g}^{\mu\nu}\,\left(f'(R_{\mu\nu}-\frac{1}{2}g_{\mu\nu}R)-\nabla_\mu\nabla_\nu f'+g_{\mu\nu}\Box f' -\frac{1}{2}g_{\mu\nu}(f'R-f) \right).
\ee
Subtracting the two variations, the higher-order terms as well as the terms proportional to the product $\psi\, G_{\mu\nu}$, $G$ being the Einstein tensor, all cancel with each other leaving only the term,
\be\label{w-actionvar}
\dot{S}_W=\frac{1}{2}\int_\mathcal{M}dv_g\, \dot{g}^{\mu\nu}\,W(\psi).
\ee
We therefore introduce the higher-order `Einstein-Legendre' (`EL') action,
\be\label{lel}
S_{\textrm{EL}}=\int_\mathcal{M} L_{\textrm{EL}}\,dv_g, \quad L_{\textrm{EL}}=R+F+X+L_{\textrm{mat}},
\ee
where  $X=(\partial\psi)^2/2$ is the kinetic term of the scalar field $\psi=f'(R)$, $L_{\textrm{mat}}$ the matter lagrangian, and the $F$-term is equal to  $F(\psi,R)$ as in Eq. (\ref{F}). Using the result (\ref{w-actionvar}), it  follows that the metric variation of the EL action (\ref{lel}), $\dot{S}_{\textrm{EL}}=0$, leads to the  field equations,
\be\label{fe}
G_{\mu\nu}=T_{\mu\nu}(\psi)+T_{\mu\nu}(\textrm{mat}),
\ee
that is to the Einstein equations with the scalar field  $\psi=f'(R)$ coupled minimally to the matter tensor $T_{\mu\nu}(\textrm{mat})$,   and  having stress tensor $T_{\mu\nu}(\psi)$  with self-interacting  potential equal to,
\be\label{pot}
V(\psi)=W(\psi)/2.
\ee
This result means that when passing to its `EL representation' (\ref{lel}), the original $f(R)$ theory  acquires a particularly simple form  given by the second-order field equations  (\ref{fe}) and obtained without any use of conformal frames. In this sense,  we say that the original $f(R)$ theory has been `scalarized'. In this way, we obtain second-order field equations through a reduction mechanism which avoids well-known issues that arise when using the Palatini method \cite{cmq}-\cite{har}.

\section{Legendre-scalarized cosmology}
In this Section, we treat homogeneous and isotropic cosmology for the quadratic theory $f(R)=R+\epsilon R^2$ in the EL representation. The potential (\ref{pot}) reads,
\be
V(\psi)=\frac{1}{8\epsilon}(\psi-1)^2,
\ee
and therefore introducing the new scalar field
\be
\chi=\frac{1}{2}(\psi-1),
\ee
the potential becomes that of a massive scalar field with an  $\epsilon$-dependent mass,
\be\label{msf}
V(\chi)=\frac{1}{2}m^2\chi^2,\quad m^2=\frac{1}{\epsilon}.
\ee
Hence, the $R+\epsilon R^2$ theory in the EL representation is described as a  massive scalar field $\chi$ in general relativity, thus simplifying the overall problem.

For an FRW universe filled with a perfect fluid with pressure $p$ and fluid density $\rho$, setting $H=\dot{a}/a$, and $\rho=\dot{\chi}^2/2+V(\chi),p=\dot{\chi}^2/2-V(\chi)$ for the massive scalar field (\ref{msf}),  the Einstein equations (\ref{fe}) are (a dot represents `$d/dt$' below),
\begin{align}\label{massf}
&\ddot{\chi}+3H\dot{\chi}+m^2\chi=0,\\
&\dot{H}+H^2=\frac{1}{6}m^2\chi^2-\frac{1}{3}\dot{\chi}^2,\\
&H^2+\frac{k}{a^2}=\frac{1}{6}\left(\dot{\chi}^2+m^2\chi^2\right).
\end{align}
Following \cite{bel,ghs}, we transform the problem in the dimensionless variables,
\be
x=\frac{1}{\sqrt{6}}\,\chi,\quad y=\frac{1}{\sqrt{6m}}\,\chi', \quad z=\frac{H}{\sqrt{m}},\quad \tau=\sqrt{m}t.
\ee
The first two equations give the three-dimensional system (a prime is `$d/d\tau$' below),
\beq
x'&=&y \label{1}\\
y'&=&-mx-3yz\label{2}\\
z'&=&mx^2-2my^2-z^2,\label{3}
\eeq
while the constraint becomes,
\be\label{con}
x^2+y^2-z^2=km^{-2}a^{-2}.
\ee
From Eq. (\ref{msf}) it follows that for $\epsilon>0$  we have the standard positive mass scalar field, whereas the case $\epsilon<0$ is  the case of  `tachyonic instability'.

We start from the \emph{massless case}, which following \cite{ghs} can be studied by putting $x=0$ in the evolution equations (\ref{1})-(\ref{3}), to get the two-dimensional system,
\beq
y'&=&-3yz\label{2'}\\
z'&=&-2my^2-z^2.\label{3'}
\eeq
\begin{figure}
     \centering
     \begin{subfigure}[b]{0.35\textwidth}
         \centering
         \includegraphics[width=\textwidth]{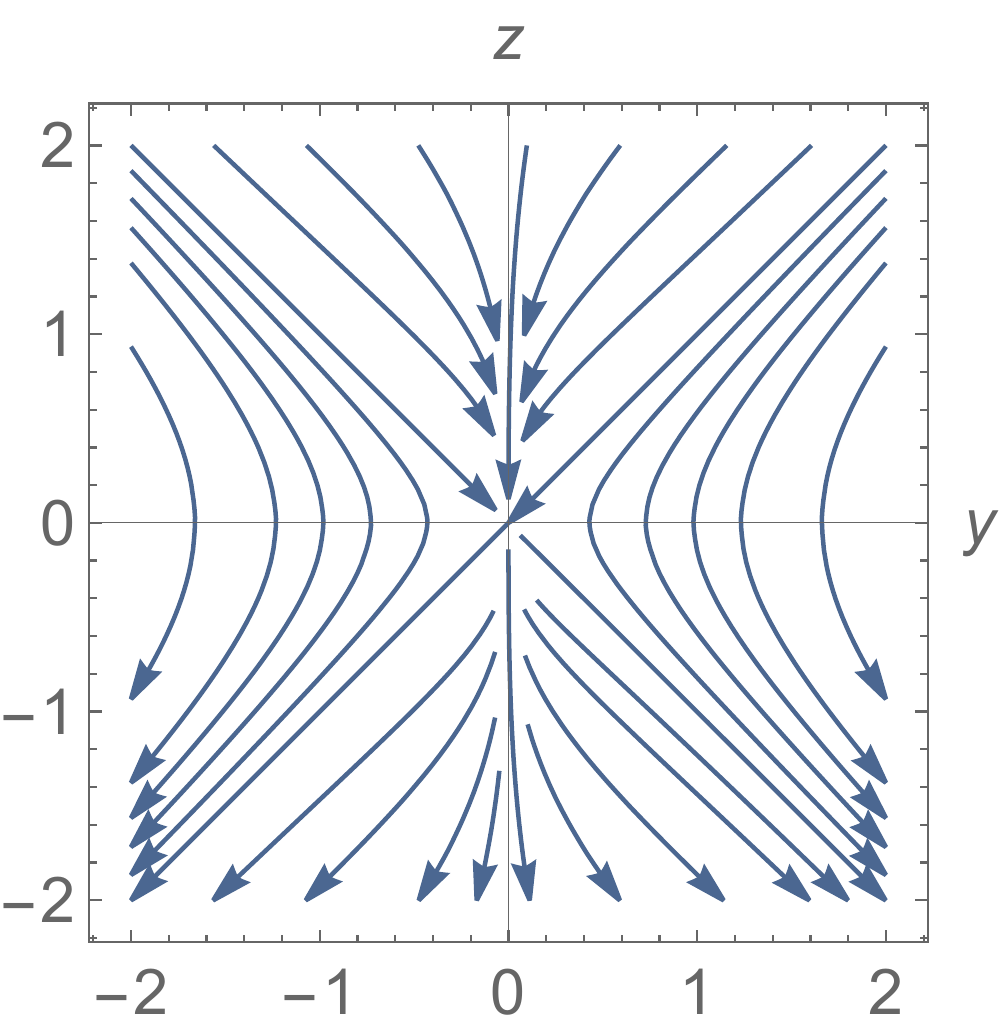}
         \caption{$m=+1$}
         \label{mu+1}
     \end{subfigure}
      %\hfill
     \begin{subfigure}[b]{0.35\textwidth}
         \centering
         \includegraphics[width=\textwidth]{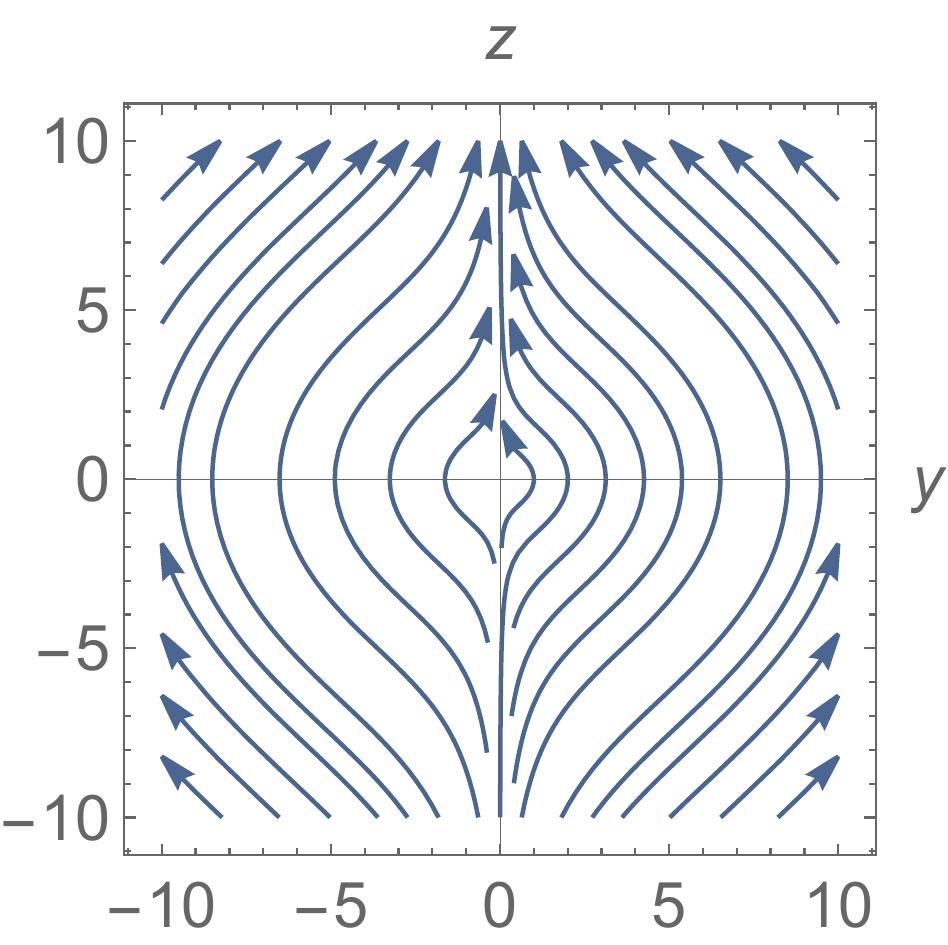}
         \caption{$m^2=-1$}
         \label{mu-1}
     \end{subfigure}
     \caption{The massless case for specific mass values. The left diagram describes the orbits for positive scalar field mass, whereas on the right we have the orbit structure for tachyonic instability.}
        \label{massless-fig}
\end{figure}
For the mass value $m^2=\pm 1$, the corresponding phase portraits  are shown in Fig. \ref{massless-fig}. The first quadrant of the portrait  of Fig. \ref{massless-fig}(a) faithfully reproduces the behaviour found in \cite{ghs} (cf. their Fig. 2), but here we consider also contracting and recollapsing universes (lower half space) as well as their `mirrors'  (left half space), all these have identical behaviour that of  an isotropic stiff fluid. The $z$-axis represents Milne universes becoming more and more empty and tending to flat space (at the origin), while those below the bisectors expand until they hit on the $z=0$ axis after which they are contracting.

The  phase diagram (b) of Fig. \ref{massless-fig} shows the behaviour of the orbits for the case of tachyonic instability, that is the case of negative mass-squared. We  further rescale the system (\ref{1})-(\ref{3}) using the linear transformation,
\be
\bar{x}=mx,\quad \bar{y}=my, \quad \bar{z}=\frac{1}{m}z,
\ee
which leads to the equivalent system,
\beq
\bar{x}'&=&\bar{y}\label{1a}\\
\bar{y}'&=&-m\bar{x}-3\bar{y}\bar{z}\label{2a}\\
\bar{z}'&=&\bar{x}^2-\frac{2}{m^2}\bar{y}^2-m^2\bar{z}^2,\label{3a}
\eeq
which incidentally implies that one cannot really remove the $m$-dependence completely from the system. Then like before we set $\bar{x}=0$ to obtain,
\beq
\bar{y}'&=&-3\bar{y}\bar{z}\label{2b}\\
\bar{z}'&=&-\frac{2}{m^2}\bar{y}^2-m^2\bar{z}^2,\label{3b}
\eeq
leading to the phase portrait of \ref{massless-fig}(b).
These orbits represent runaway universes evolving from infinite contraction to infinite expansion avoiding flat space.
When $k=0$ the situation simplifies because from Eq. (\ref{con}) the constraint becomes $z^2=x^2+y^2$. This means that the motion takes place on the surface of a cone with vertex at the origin. Setting $m=1$, the first two equations (\ref{1}), (\ref{2}) become
\beq
x'&=&y \label{1c}\\
y'&=&-x-3y\sqrt{x^2+y^2}, \label{2c}
\eeq
while using these two equations,
\be
z'=-3y^2,\label{3c}
\ee
so that $z$ continuously decreases to zero. It is not difficult to see that the function $g(x,y)=(x^2+y^2)/2$ is a Liapunov function of the system. (For other cosmological applications of this technique, see \cite{jo}.) This implies that the state at the origin is globally asymptotically stable in this case.

\begin{figure}[tbh]
\begin{center}
\includegraphics{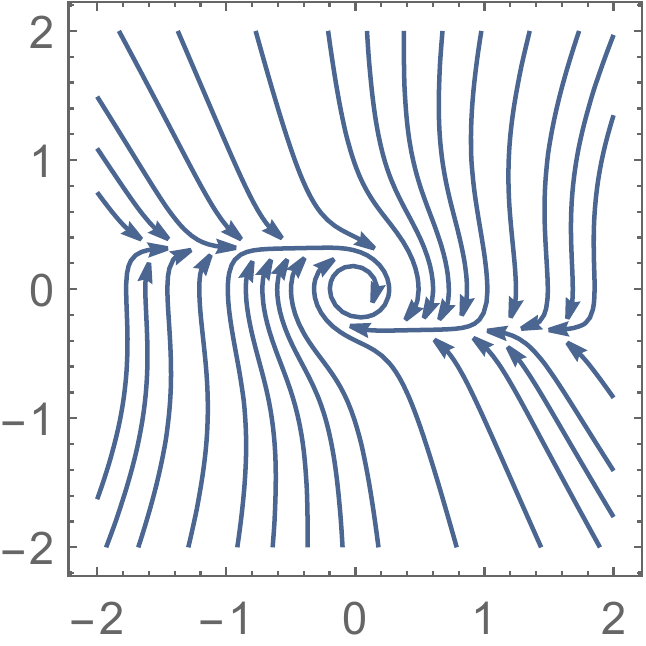}
\end{center}
\caption{Motion on the cone for a massive scalar field ($\epsilon>0$).}%
\label{flat1}
\end{figure}

Passing to polar coordinates $(r=z,\theta)$, Eqns. (\ref{1c}), (\ref{2c}) become,
\beq
r'&=&-3r^2\sin^2\theta, \label{1d}\\
\theta'&=&-1-3r\sin\theta\cos\theta, \label{2d}
\eeq
for the motion on the cone. We then observe that for $r>0$, $r'<0$, which implies that orbits spiral inwards, making the origin a stable focus. This dynamics is captured in Fig. \ref{flat1},
to be compared with Fig. 1a of Ref. \cite{bel}. The results are qualitative the same.

These results allow for a comment with regard to inflation in the $R+\epsilon R^2$ theory, in  its present Legendre-scalarized version (\ref{fe}) with the scalar field potential given by  (\ref{msf}).
For the system (\ref{1d}), (\ref{2d}), the center manifold in this case was calculated by A. Rendall and reads (\cite{ren}, Sec. 6),
\be\label{ren1}
u'=\frac{1}{3}u^3+\mathcal{O}^4,
\ee
where $u=1-\rho, \rho=r/(1+r)$ (cf. also \cite{ghs}, Eqn. (3.24) for an analogous result).
Since $\tau=\sqrt{m} t$ in our variables, this implies that $z\equiv H/\sqrt{m}\sim t$, and so we find that
\be\label{inf}
a(t)\sim e^{m^2t^2},
\ee
with $m=1/\sqrt{\epsilon}$ from Eq. (\ref{msf}).
Now it follows from general theorems (cf. \cite{wig} and refs. therein) that  the center manifold for the system (\ref{1})-(\ref{con}) in the flat case ($k=0$) and \emph{for any $m$ near $m=1$}, differs from that of the flat case with $m=1$ (that is from (\ref{ren1})) only by transcendentally small terms, and so we do not need to calculate the former anew.
Therefore we have shown that inflation proceeds as usual in the Legendre scalarized version of the
theory since the result (\ref{inf}) is valid for it as it did for the original $R + \epsilon R^2$ version of the theory, cf.  \cite{ba-ot,star,mms,cf-93}.

\section{Discussion}
The Einstein-Legendre approach introduced and studied in this paper recasts  $f(R)$ theory in the form `general relativity plus a self-interacting  scalar field' \emph{without} introducing frames and conformal transformations, but staying with the same metric  and Legendre-scalarizing the action.  This avoids a number of shortcomings of the standard conformal equivalence such as the physicality issue, and the problem of the presence of nonminimal couplings of the field $\psi$ to matter which arise when one passes to a conformally related frame, while at the same time reduces the  equations to second order. The present approach is also simpler in that \emph{polynomial} potentials (instead of exponential ones) result without conformal transformations, compare e.g., the case of the quadratic theory potential (\ref{msf}) studied here to the polynomial potential that arises in the classic  result of Whitt \cite{whitt}. The developments of Section 3 of this paper hopefully highlight this point more clearly.

One may extend the Legendre scalarization reasoning advanced presently to  other interesting choices of modified gravity. For example, for a theory of the form  $f(R, \textrm{Ric}^2)\equiv f(x,y)$, we may set $\phi=\dot{f}=\pa f/\pa x, \psi=f'=\pa f/\pa y$, and the Legendre transform in this case will be,
$
W:(\phi,\psi,f(x,y))\mapsto(\phi,\psi,W(\phi,\psi)),
$
where,
$
W(\phi,\psi)=\phi x+\psi y-f(x,y).
$
In this case  Legendre scalarization  will be realized in the form of  the following Einstein-Ricci-Legendre lagrangian,
$
R+W+X_\phi +X_\psi+L_{\textrm{mat}},
$
and we may further proceed as above. Other interesting terms of higher-order may be scalarized by using higher-dimensional Legendre transforms.

\section*{Acknowledgments}
The research of S.C.  was funded by RUDN university  scientific project number FSSF-2023-0003. JPM thanks the Funda\c c\~ao para a Ci\^encia e Tecnologia (FCT) for the financial support of the grants EXPL/FIS-AST/1368/2021, PTDC/FIS-AST/0054/2021, UIDB/04434/2020, UIDP/04434/2020, and CERN/FIS-PAR/0037/2019.

\end{document}